# Shannon's entropy revisited

Sergei Viznyuk

**Abstract**

I consider the effect of a finite sample size on the entropy of a sample of independent events. I propose formula for entropy which satisfies Shannon's axioms, and which reduces to Shannon's entropy when sample size is infinite. I discuss the physical meaning of the difference between two formulas, including some practical implications, such as maximum achievable channel utilization, and minimum achievable communication protocol overhead, for a given message size.

The classic formula (Shannon, 1948)
$$H_S = - \sum_i p_i \ln p_i \qquad (1)$$

is accepted as a measure of entropy for a sample of independent events. Other measures, such as (Renyi, 1960) and (Tsallis, 1988) entropies, and more (Sharma B.D., 1975; Masi, 2005; Gorban, 2002) have been proposed for situations when events are correlated, as e.g. in systems with long-range interactions. These alternative measures, generally, are expected to have properties (Wehrl, 1978), different from Shannon's entropy: non-additivity, convexity, etc., while converging to (1) when correlations disappear. The underlying assumption is that when events are independent formula (1) still gives the correct measure.

Here I take a second look at the original premise that formula (1) and its equivalents in statistical (Landau, 1980) and quantum (Neumann, 1955) mechanics are the correct measures of entropy in the absence of correlations. I show that if the sample size is finite there is an inherently smaller amount of information which can be encoded per event, than given by Shannon's formula. I propose a formula for entropy which takes into account the sample size and which reduces to (1) when sample size is infinite.

Consider $N$ independent events, with $M$ possible outcomes for each event, having probabilities $p_i$, $1 \leq i \leq M$. Independence of events implies only the frequencies $\{n_i\}$ of events describe the sample, while the order of events is irrelevant. The probability density function for the sample is then given by multinomial distribution:

$$P(\{n_i\}; N, \{p_i\}) = N! \prod_i \frac{p_i^{n_i}}{n_i!} = \Omega_{\{n_i\}} \cdot \prod_i p_i^{n_i} \qquad (2)$$

The factor $\Omega_{\{n_i\}}$ in (2) is *statistical weight*:
$$\Omega_{\{n_i\}} = N! \prod_i \frac{1}{n_i!} \qquad (3)$$

The starting point is the following definition of entropy (Huang, Statistical Mechanics, 1987):
$$H_\Omega = \ln \Omega \qquad (4)$$

With (3) I rewrite (4) as:
$$H_\Omega(\{n_i\}; N, M) = \frac{1}{N}\left[\ln \Gamma(N+1) - \sum_{i=1}^M \ln \Gamma(n_i + 1)\right] \qquad (5)$$

, where $\Gamma$ is *gamma* function. In (5) I divided $H_\Omega$ by the number of events $N$ to obtain per-event entropy, instead of the total entropy of the sample given by (4). I'm interested in the expression for entropy in *equilibrium*, i.e. the expression which maximizes probability (2). The equilibrium is achieved when $n_i = Np_i \ \forall \ i$. Thus the equilibrium entropy $H_\Omega$ is (Viznyuk, 2014):

$$H_\Omega(\{p_i\}; N, M) = \frac{1}{N}\left[\ln \Gamma(N+1) - \sum_{i=1}^{M} \ln \Gamma(Np_i + 1)\right] \quad (6)$$

Using Stirling's approximation for factorials, the difference between (6) and (1) is

$$(H_S - H_\Omega)|_{N \to \infty} = \frac{1}{2N}\left[(M-1)\ln 2\pi N + \sum_{i=1}^{M} \ln p_i\right] \to 0 \ when \ N \to \infty \quad (7)$$

It shows Shannon's entropy $H_S$ is a limit case of $H_\Omega$ when $N \to \infty$, i.e. in *thermodynamic limit* (Huang, 2001) approximation, except when $H_S$ diverges (Baccetti, 2013), in which case it is worth noting the following asymptotic behavior:

$$H_\Omega = \gamma + \frac{\ln \Gamma(N+1)}{N} \quad , when \ Np_i \to 0 \ \forall \ i \quad (8)$$

$$H_S = \ln M \quad , when \ p_i = 1/M \ \forall \ i \quad (9)$$

, where $\gamma \cong 0.5772156649$ is *Euler-Mascheroni* constant. The expressions (1),(6),(8),(9) are for the *per-event* entropy, in units of *nats*. To find *per-event* entropy in units of *bits*, these expressions have to be divided by $\ln 2$. If each event is encoded as a sequence of symbols from some hypothetical alphabet **B**, one can invent a new unit of entropy and call it, say, *bean*. The occurrence of one symbol from **B** is one *bean*. To find *per-bean* entropy in units of *beans*, one has to divide expressions (1),(6),(8),(9) by $\ln M$. Remarkably, the result does not depend on alphabet **B**. The values of $H_S$ and $H_\Omega$ entropies in *beans per bean* never exceed 1, even as the *per-event* entropy $H_S$ in (1) and (9) may diverge when $M \to \infty$. I elaborate more on this below.

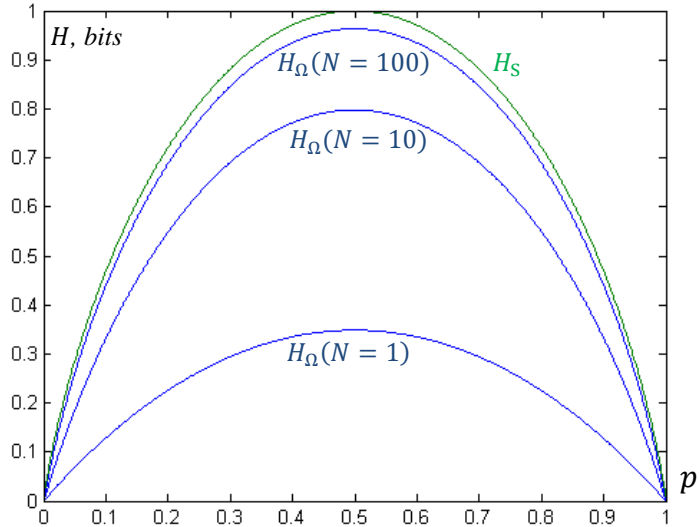

Figure 1
Convergence of $H_\Omega$ entropy to $H_S$ as sample size $N$ increases, for the case of $M = 2$ possible event outcomes

It is easy to see $H_\Omega$ entropy in (6) satisfies all three so-called Shannon's axioms (Petz, 2001):
    (a) Continuity: $H(p, 1-p)$ is a continuous function of $p$.
    (b) Symmetry: $H(p_1, p_2, \ldots, p_n)$ is a symmetric function of its variables.

(c) Recursion: For every $0 \leq \lambda < 1$ the following recursion holds:
$$H(p_1, p_2, \ldots, p_{n-1}, \lambda p_n, (1-\lambda)p_n) = H(p_1, p_2, \ldots, p_n) + p_n \cdot H(\lambda, (1-\lambda))$$

Properties (a-b) are obviously satisfied. To prove property (c) I break the outcome with index $i = M$ and probability $p_M$ in (6) into 2 sub-outcomes having relative probabilities $\lambda$ and $(1-\lambda)$, and absolute probabilities $\lambda p_M$ and $(\lambda - 1)p_M$. The new number of possible outcomes is $M + 1$.
I re-write (6) for the $M + 1$ number of outcomes as: $H_\Omega(\{p_i\}; N, M+1) =$

$$\frac{1}{N}\left[\ln \Gamma(N+1) - \sum_{i=1}^{M-1} \ln \Gamma(Np_i + 1) - \ln \Gamma(N\lambda p_M + 1) - \ln \Gamma(N(\lambda - 1)p_M + 1)\right] =$$

$$\frac{1}{N}\left[\ln \Gamma(N+1) - \sum_{i=1}^{M} \ln \Gamma(Np_i + 1)\right] + \tag{10}$$

$$p_M \cdot \frac{1}{Np_M}[\ln \Gamma(Np_M + 1) - \ln \Gamma(N\lambda p_M + 1) - \ln \Gamma(N(\lambda - 1)p_M + 1)] =$$

$$H_\Omega(\{p_i\}; N, M) + p_M \cdot H_\Omega(\lambda, (1-\lambda); Np_M, 2)$$

A general recursive property can be derived by combining possible $M$ event outcomes into $K$ non-overlapping groups, with $N_k$ events having $M_k$ possible outcomes in group $k$: $1 \leq k \leq K$; $\sum_k N_k = N$; $\sum_k M_k = M$; $NP_k = N_k$. Then (6) can be rewritten as the sum of *coarse-grained* entropy $H_\Omega(\{P_k\}; N, K)$ and the weighted sum of individual *outcome group* entropy values (Viznyuk, 2014):

$$H_\Omega(\{p_i\}; N, M) = \frac{1}{N}\left[\ln \Gamma(N+1) - \sum_{k=1}^{K} \ln \Gamma(NP_k + 1)\right] +$$

$$\sum_{k=1}^{K} P_k \cdot \frac{1}{N_k}\left[\ln \Gamma(N_k + 1) - \sum_{i \in M_k} \ln \Gamma\left(N_k \frac{p_i}{P_k} + 1\right)\right] \tag{11}$$

$$= H_\Omega(\{P_k\}; N, K) + \sum_{k=1}^{K} P_k H_\Omega\left(\left\{\frac{p_{i \in M_k}}{P_k}\right\}; N_k, M_k\right)$$

Shannon has attached a practical significance to entropy in the form of his *Fundamental Theorem for a Noiseless Channel*:

"*Theorem 9:* Let a source have entropy $H$ (bits per symbol) and a channel have a capacity $C$ (bits per second). Then it is possible to encode the output of the source in such a way as to transmit at the average rate $(C/H) - \epsilon$ symbols per second over the channel where $\epsilon$ is arbitrarily small. It is not possible to transmit at an average rate greater than $C/H$."

I restate Shannon's theorem using dimensionless *beans per bean* units of entropy as follows:

If $H$ is entropy in the channel in units of *beans per bean*, the channel utilization is $H$.

With this understanding, $H_\Omega$ entropy derived above acquires the following meaning:

1. $H_\Omega(\{p_i\}; N)$ *beans per bean* is the maximum channel utilization which can be achieved when transmitting messages of $N$ symbols long with symbol probability distribution $\{p_i\}$.
2. $\left(H_S(\{p_i\}) - H_\Omega(\{p_i\}; N)\right)$ *beans per bean* is the minimum protocol overhead which can be achieved when transmitting messages of $N$ symbols long with symbol probability distribution $\{p_i\}$.

Consider a simple example when a source sends a message of $N$ bits to a recipient. At the minimum the source has to communicate the size of the message recipient should expect. This information will take additional $\log_2 N$ bits. Thus, the total size which has to be transmitted is $N + \log_2 N$. Per a bit of message received at destination $\frac{N}{N+\log_2 N}$ is the payload and $\frac{\log_2 N}{N+\log_2 N}$ is the protocol overhead. According to the meaning of $H_\Omega$ entropy stated above, the maximum achievable payload is (in bits per bit):

$$H_\Omega = \frac{1}{N \cdot \ln 2}\left[\ln \Gamma(N+1) - 2 \cdot \ln \Gamma\left(\frac{N}{2}+1\right)\right] \quad (12)$$

, and the minimum achievable protocol overhead is: $\quad 1 - H_\Omega \quad (13)$

For a message of $N = 256$ bits long, from (12): $H_\Omega = 0.9831 > \frac{256}{256+\log_2 256} = 0.9697$

For a message of $N = 16$ bits long: $H_\Omega = 0.8532 > \frac{16}{16+\log_2 16} = 0.8$

As another example, for the Ethernet frame payload size of 1500 bytes = 12000 bits I get $1 - H_\Omega = 5.9176 \cdot 10^{-4}$ bits per bit, assuming ideal compression with $H_S = 1$. The Ethernet header size is 26 bytes=208 bits which gives the real protocol overhead of 0.017 bits per bit.

Shannon's entropy (1) diverges for some probability distributions (Baccetti, 2013). Divergence of $H_S$ happens because for those distributions it takes, in average, an infinite number of symbols in any alphabet to encode the event stream. On the other hand, entropy $H_\Omega(\{p_i\}; N)$ never diverges because it is always possible to encode $N < \infty$ events using finite number of symbols from some coding scheme.

Let $\{p_i > 0\}, 1 \leq i < \infty$ be an arbitrary probability distribution. Define a partial sum $Z(M) = \sum_{i=1}^{M} p_i$, and probability distribution $\{q_k\} = \{p_k/Z(M)\}, 1 \leq k \leq M$. Let $H_S(\{q_k\})$ be Shannon's entropy of $\{q_k\}$ distribution in *beans per bean* units, i.e. divided by $\ln M$:

$$H_S(\{q_k\}) = -\frac{1}{\ln M}\sum_{k=1}^{M} q_k \ln q_k \quad (14)$$

As $M \to \infty$ in (14) the value of $H_S(\{q_k\})$ converges to *beans per bean* entropy of $\{p_i\}$ distribution. It is straightforward to prove (14) never exceeds 1 for any probability distribution, and $H_S = 1$ *beans per bean* if and only if $p_i = 1/M \ \forall \ i$.